\documentclass[a4paper]{IEEEtran}
\usepackage{color}
\usepackage{leftidx}
\usepackage{cite}
\usepackage{microtype}
\usepackage{cleveref}
\usepackage{graphicx,amssymb,amstext,amsmath}
\usepackage[ruled,vlined,lined,ruled,linesnumbered]{algorithm2e}

\usepackage{float}
\usepackage{subfigure}
\usepackage{emptypage}

\begin{document}
\title{Computation Rate Maximization in Wireless Powered MEC with Spread Spectrum Multiple Access }
\author{Yuegui Chen, Suzhi Bi, Xian Li, Xiaohui Lin, and Hui Wang\\
 College of Electronic and Information Engineering, Shenzhen University, Shenzhen, China.\\
  E-mail: chenyuegui2017@email.szu.edu.cn, \{bsz, xianli, xhlin, wanghsz\}@szu.edu.cn.}

\vspace{-2ex}

\maketitle
\begin{abstract}
 The integration of mobile edge computing (MEC) and wireless power transfer (WPT) technologies has recently emerged as  an effective solution for extending battery life and increasing the computing power of wireless devices. In this paper, we study the resource allocation problem of a multi-user wireless powered MEC system, where the users share the wireless channel via direct sequence code division multiple access (DS-CDMA). In particular, we are interested in jointly optimizing the task offloading decisions and resource allocation, to maximize the weighted sum computation rate of all the users in the network. The optimization problem is formulated as a mixed integer non-linear programming (MINLP). For a given offloading user set, we implement an efficient Fractional Programming (FP) approach to mitigate the multi-user interference in the uplink task offloading. On top of that, we  then  propose a Stochastic Local Search algorithm to optimize the offloading decisions. Simulation results show that the proposed method can effectively enhance the computing performance of a wireless powered MEC with spread spectrum multiple access compared to other representative benchmark methods.
\end{abstract}


\vspace{-2ex}
\IEEEpeerreviewmaketitle

\section{Introduction}

With the explosive development of the Internet of things (IoT), new applications have emerged with advanced features that require consistent high-performance computing service and low processing latency, such as autonomous driving, smart appliances, virtual reality and remote surgery. New computing performance requirements have also spawned a new computing paradigm named Mobile Edge Computing (MEC), which deploys cost-efficient servers  within the radio access network, e.g., at WiFi access points (APs) or the cellular radio network controller site, to reduce the backhaul transmission delay in traditional cloud computing system \cite{23:}. There has been extensive studies on optimizing the MEC system computing  performance by jointing optimizing task offloading decisions and system resource allocation \cite{11:,16:,17:}.  However, the performance of conventional  MEC system  is limited by the finite battery life of WDs, especially for size-constrained IoT devices.  Recently, radio-frequency (RF) signal based wireless power transfer (WPT) provides a viable solution that can charge the battery continuously in distance without interrupting its normal operation \cite{2:,3:}.  There have been some recent studies on integrating WPT to conventional MEC systems \cite{20:,19:,202:}. For instance, \cite{20:} considers a multi-user wireless powered MEC network and maximizes the weighted computation rate of all the users by jointly optimizing the task offloading decisions and the system transmission time allocation.  The work is later extended in \cite{19:} to use a deep reenforcement learning approach for optimizing the offloading decisions.

The existing studies have considered many multiple access schemes for the task-offloading users to share the common spectrum, such as  TDMA (e.g., in \cite{17:}, \cite{20:}),  FDMA (e.g., in \cite{7:}), OFDMA (e.g., in \cite{15:})  and  non-orthogonal multiple access (NOMA) (e.g., in  \cite{9:}).  In pracitce,  spread spectrum multiple access  (such as DS-CDMA) is a standardized multiple access method widely used in  IoT systems, such as in LoRaWAN and SigFox \cite{27:}, due to its simplicity in hosting massive number of simultaneous connections. Nevertheless, to the authors' best knowledge, it lacks relevant research in prior works on applying spread spectrum multiple access to wireless powered MEC. In this paper, we consider a multiuser wireless-powered MEC system using DS-CDMA to coordinate the task offloading of the users.  Our main goal is to maximize the weighted sum of computation rates across all the users. The main contributions of this paper are as follows:
\begin{enumerate}
  \item
  We study the computation rate maximization problem in a wireless powered MEC system where the users perform task offloading based on DS-CDMA. The problem is formulated as a mixed integer non-linear programming (MINLP) that jointly optimizes the time allocation on WPT and task offloading, the binary offloading decisions of the WDs, and the transmit power control of the offloading WDs.

  \item
  To tackle the problem, we first consider the case that the set of offloading WDs is given. The remaining optimization on time allocation and multi-user power control, however, is still a hard non-convex problem. For a given WPT time, we adopt a fractional programming(FP) approach to maximize the weighted sum computation rate of the users in the interference channel. Then, the optimal time allocation can be easily obtained through a simple one-dimension search method, e.g., golden-section search.
  \item
  On top of the WPT time allocation and power control method, we further propose an integer Stochastic Local Search  method to optimize the combinatorial computation offloading decisions. The method is based on iterative update of user binary offloading decision, and shown to converge to a good solution within limited iterations.

\end{enumerate}

\section{System Model}
\subsection{ Network Model }
As shown in Fig. 1, we consider a multiuser wireless-powered MEC system with $N$ users and one access point (AP) that is integrated with a RF energy transmitter and an MEC server. Each of the device is equipped with a single antenna. The AP broadcasts RF energy while all users harvest energy and charge the battery with the received energy. In this paper, we consider the binary computation offloading policy\cite{23:}, where the task is indivisible, such that a WD either operates in offloading mode  (mode 1, like WD1 and WD2 in Fig. 1) or local computing mode (mode 0, like WD3). Let  $\mathcal{N}_1$ and $\mathcal{N}_0$ denote the two non-overlapping index sets of WDs that operate in mode-1 and mode-0, respectively, where $ \mathcal{N}=\mathcal{N}_1\;  {\cup} \; \mathcal{N}_0 =\{1,...,N\}$ denotes the set of all the users.  To avoid co-channel interference, the downlink WPT and the uplink wireless communication (for offloading) operate in a time-division-multiplexing (TDD) manner. The mode-1 users offload their task data in  the  uplink to the MEC server simultaneously using CDMA. As shown in Fig.1, within each system time frame of duration $\emph{T}$, the transmission time  is divided into three phases: the first  phase ${\alpha}T$  is used for the AP broadcasting RF energy to the WDs, where $\alpha \in (0,1)$, the second phase is for the mode-1 users offloading the task data, and  the AP computes and returns the computation results back to the users in the last phase.  Because of the much stronger computation capability and transmit power of the AP, as well as the relatively short computation result, we neglect the time consumed on task computing and result downloading (as in \cite{20:},\cite{7:}). Therefore, the second phase occupies $(1-\alpha)T$ amount of time.

During the first period ${\alpha}T$, HAP broadcasts RF power to all WDs in the downlink with fixed transmit power $P_0$, the energy collected by the $i$-th user is expressed as follows
\footnote{We consider a linear energy harvesting model in this paper, as it is shown that the non-linear saturation effect of a single energy harvesting circuit can be effectively rectified by using multiple energy harvesting circuits concatenated in parallel, resulting in a sufficiently large linear conversion region in practice \cite{2019:Kang}. }:
\begin{equation}
\label{energy}
\small
E_i={\nu}{h_i}P_0{\alpha}T, i=1,2,...,N,
\end{equation}
where $h_i$  denotes the channel power gain from the AP to the $i$-th WD$s$, which is assumed equal for both the uplink and downlink for simplicity.  $\nu\in(0,1)$ is the energy harvesting efficiency. After WET, mode-1 users offload their tasks for edge computation for the rest of the time frame.
\begin{figure}
  \centering
   \begin{center}
      \includegraphics[width=0.45\textwidth]{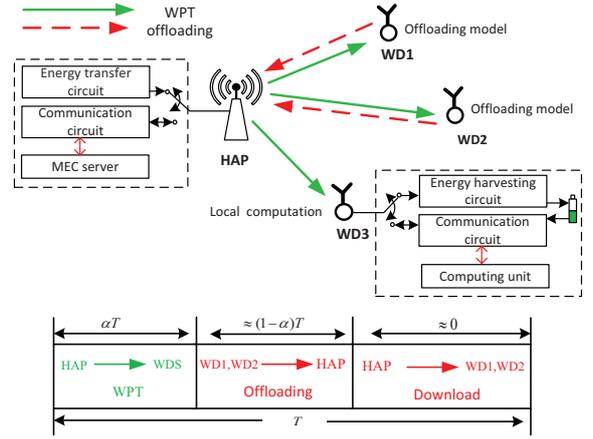}
   \end{center}
  \caption{The model for 3-user WPT-MEC system with binary computation offloading and system time allocation}
  \label{Fig.1}
\end{figure}
\subsection{Offloading Mode }
Due to the TDD circuit constraint, a mode-1 user starts offloading its task data for edge execution using DS-CDMA after harvesting RF energy. Suppose that a pseudo-random code of processing gain $G$ is used as the spreading sequence, the maximum amount of information offloaded to the edge by the $i$-th user is [14, Sec.  4]
\begin{small}
\begin{align*}
&I_{off,i}(\mathbf{P},\alpha)\\&={\frac{B{(1-\alpha)T}}{G}}\log_{2}\left(1+\frac{G{P_i}{h_i}}{\sum_{n\in \mathcal{N}_1, n \neq i,}{P_n}{h_n}+N_0{B}}\right),\tag{2}
\end{align*}
\end{small}
$\forall i\in \mathcal{N}_1,$ where $P_i$ denotes the transmit power of the $i$-th user. $B$ and $N_0$ denote the channel bandwidth and the receiver noise power. $\sum_{n\in \mathcal{N}_1, n \neq i,}{P_n}{h_n}$ represents the  interference caused by the other offloading users to the  $i$-th user due to the imperfect spreading code. Accordingly, the task offloading rate is
\begin{align*}
r_{off, i} (\mathbf{P},\alpha) = I_{off,i} /T,\ \ \forall i\in \mathcal{N}_1. \tag{3}
\end{align*}
As we neglect the edge computation time and result downloading time, $r_{off,i}$ equals to the number of processed raw data per second. Under the energy harvesting constrain, the transmit power $P_i$ of $i$-th user is limited by ${P_i}{(1-\alpha)}T\leq{E_i}+e_i$, where $e_i$ denotes the residual energy left  from the previous time frame and is assumed zero for simplicity and without loss of generality in the following analysis. Equivalently, we have
 \begin{align*}
P_i \leq \max\left\{ {\frac{{\nu}h_i{\alpha}{P_0}}{1-{\alpha}}}, q_{max}\right\}\triangleq P_{max,i},\tag{4}
\end{align*}
where $q_{max}$ denotes the maximum transmission power and $P_{max,i}$ denotes the
maximum allowable power of the $i$-th WD due to the energy harvesting constraint (1). Since the value of $h_i$ and $e_i$ for each WD are different, the corresponding maximum power $P_{max,i}$ is also different between different users in general.
\subsection{Local Computing Mode}
Because energy harvesting and task computation use separate circuits at the user device,  a mode-$0$ user  can harvest RF energy and compute its task at the same time, which indicates that  it can compute for the entire duration of $T$.\footnote{Here, we consider that each WD has sufficient initial energy to perform local computation at the beginning of a time frame (such as in \cite{20:}).} Let $C_i > 0 $ denote the number of CPU cycles required to process one bit of raw data, which is assumed equal for all the WDs. Let  $f_i$ represent the CPU frequency (cycles per second) for processing the task of the $i$-th user.  It is shown in \cite{20:} that an energy-constrained user should compute throughout the time frame to maximize its computation rate, i.e., the number of information bits processed within a unit time. Accordingly, the energy on computation is constrained  as\cite{20:}
\begin{align*}
k_i f_i^3 T\leq E_i, \forall i \in \mathcal{N}_0,\tag{5}
\end{align*}
where $k_i$ denotes the computation energy efficiency coefficient. By exhausting the harvested energy, we have $f^{*}_i=(\frac{E_i}{k_i{T}})^{\frac{1}{3}}$. By substituting (1) to (5), the maximum local computation rate of the $i$-th user is
\begin{small}
\begin{align*}
{r}^{*}_{loc,i}\left(\alpha\right)=\frac{f_i^*}{C_i}=  \frac{\left(\nu{P_0}\right)^{\frac{1}{3}}}{C_i}{\left(\frac{h_i}{k_i}\right)^{\frac{1}{3}}}{{\alpha}^{\frac{1}{3}}}={\eta}%
{\left(\frac{{{h_i}}}{{{k_i}}}\right)^{\frac{1}{3}}}{\alpha ^{\frac{1}{3}}}, i\in{\mathcal{N}_0},\tag{6}
\end{align*}
\end{small}
where $\eta=\frac{\left(\nu{P_0}\right)^{\frac{1}{3}}}{C_i}$ is a fixed parameter.
\subsection{Problem Formulation }
In this article, our main purpose is to maximize the weighted sum  computation  rate of
all  users in a time frame, which reflects the data processing capability of the considered wireless powered MEC network. Based on  the analysis above, the computation rate of each user is related to its computation offloading  decisions. Let $w_i>0$ represent the positive weight assigned to the $i$-th user. We denote the offloading  strategy as $\mathbf{x}=\{x_i,{\forall}{i}\in{\mathcal{N}}\}$, where $x_i=1$ denotes that the $i$-th user offloads  its computation  task and $x_i=0$ denotes that it  computes  the computation task locally. Accordingly, the weighted sum computation rate of the wireless powered MEC network in a time frame is denoted as
\begin{align*}
F\left(\mathbf{x},\mathbf{P},\alpha\right)=\sum\limits_{i = 1}^N {{w_i}} \left(1 - {x_i}\right)r_{loc,i}^*\left(\alpha\right) + {w_i}{x_i}{r_{off,i}\left(\mathbf{P},\alpha\right)}.\tag{7}
  \end{align*}

Mathematically, the rate maximization problem is formulated as follows:
\begin{align*}
(\rm{P1}): F^{*}=& \max_{\mathbf{x},\mathbf{P},{\alpha}} \quad F(\mathbf{x},\mathbf{P},\alpha)\tag{8a}
    \\
    &\text{subject  to}\;\\&
    P_i\leq \max\left\{ {\frac{{\nu}h_i{\alpha}{P_0}}{1-{\alpha}}}, q_{max}\right\},\ \ \forall i\in{\mathcal{N}_1},\tag{8b}
    \\
    & 0{\leq}\alpha{\leq}1,x_i\in\{0,1\},  i = 1,\cdots, N.\tag{8c}
  \end{align*}
Problem (P1) is highly non-convex  and hard to solve for optimum. Firstly, it contains the combinatorial mode selection binary variables. Secondly, even if the computation modes are given, the remaining multi-user power control in interference channel is a non-convex hard problem.  Intuitively, it is not optimal for all users to compute locally or to offload all their tasks for edge execution. On the one hand, by letting all users to  offload their tasks will result in severely high interference levels, thus reducing the total  task offloading rate. It is therefore important to select a proper set of mode-1 users to maximize the computation rate. In the following section, we propose an efficient algorithm to optimize (P1).
\section{ Proposed Algorithm to (P1) }
  We decompose the original problem into two subproblems: computation resource allocation  on $\{\alpha,P\}$ for a fixed $\mathbf{x}$ and offloading decision optimization to find the optimal $\mathbf{x}$. In this section, we show how to solve the two subproblems respectively.
\subsection{ Optimal Time Allocation and Power Control given $\mathbf{x}$ }
To begin with, we assume $\mathbf{x}$ (or equivalently $\mathcal{N}_0$) and $\alpha$ are given, the problem reduces to the following power control problem
\begin{align*}
F\left(\mathcal{N}_0,\alpha \right) = \max_{\mathbf{P}} F\left(\mathcal{N}_0,\alpha,\mathbf{P}\right). \tag{9}
\end{align*}
Notice once $\mathcal{N}_0$ and $\alpha$ are given, the computation rate of each mode-0 user are automatically obtained from (6). Besides, the transmit power constraint $P_{max,i}$ are  fixed value for each mode-1 user from (4). Accordingly, we only need to optimize the weighted sum rate of all mode-1 users in (P1) by controlling each user's transmit power under respective maximum power constraint. The objective of (P1) is reduced to
\begin{align*}
R(\mathbf{P})=\sum_{k\in{\mathcal{N}_1}}{w_k}r_{off,k}(\mathbf{P}),\tag{10}
\end{align*}
where $\mathbf{P}= \left\{P_k, \forall k\in \mathcal{N}_1 \right\}$ denotes the transmission power vector of mode-1 users. We aim to find the optimal power allocation ${\mathbf{P}}^*$ that maximizes the computation rates of the mode-1 users, which is formulated as
\begin{align*}
   &\mathop {{\mathop{\rm maximize}\nolimits} }\limits_{\bf{P}} \quad R({\bf{P}})\tag{11a}
   \\
    &\text{subject  to} \quad \quad  P_i\leq P_{max,i},\ \ \forall i\in{\mathcal{N}_1}.\; \tag{11b}
\end{align*}
 Note  that (11) is a non-convex problem, which corresponds to a classic multi-user power control problem in an uplink wireless communication system. In the following subsection, we implement a direct FP approach to find a near-optimal solution.
 It is shown in \cite{25:} that the direct FP approach is effective in control the transmit power of multiple users to maximize the weighted sum throughput performance in the uplink interference channel. For simplicity of illustration, we denote $M =|\mathcal{N}_1|$. We introduce auxiliary variables ${y_i}'s$ and apply  the quadratic transform technique to each SINR term, which leads to the following reformulation
\begin{align*}
  & \mathop {{\mathop{\rm maximize}\nolimits} }\limits_{{\bf{P}},{\bf{y}}} \quad f({\bf{P}},{\bf{y}})\tag{12a}
   \\
    &\text{subject  to} \quad \quad  P_i\leq P_{max,i},\ \ \forall i\in{\mathcal{N}_1}\; \tag{12b}
    \\
    &\quad \quad \quad \quad  \quad \quad  y_i\in R,\ \ \forall i\in{\mathcal{N}_1}\; \tag{12c}
\end{align*}
where the new objective $f$ is

 \begin{small}
 \begin{align*}
    &f(\mathbf{P},\mathbf{y})=
   \sum\limits_{i = 1}^{M} {{w_i}}{\frac{B{(1-\alpha)T}}{G}}\log_{2}\bigg[1+2y_i\sqrt{G{P_i}{h_i}}
    \\&
  \quad \quad \quad \quad  \quad \quad \quad \quad \quad \quad  -{y_i}^{2}\left({\sum_{n\in \mathcal{N}_1, n \neq i,}{P_n}{h_n}+N_0{B}}\right)\bigg]. \tag{13}
\end{align*}
\end{small}
When $\mathbf{y} $ is fixed, the quadratic transform (13) is concave in $\mathbf{P}$. Therefore, the optimal $\mathbf{P}$ can  be efficiently obtained through conventional  convex optimization algorithms, e.g., interior point method.

On the other hand, when $\mathbf{P}$ is held fixed, the optimal $y_i$ can be found in closed
form as
 \begin{align*}
 {y_i}^{*}=\frac{\sqrt{G{P_i}{h_i}}}{{\sum_{n\in \mathcal{N}_1, n \neq i,}{P_n}{h_n}+N_0{B}}}, \forall i \in \mathcal{N}_1.\tag{14}
 \end{align*}

 Therefore, we can  optimize $y_i$ and $P_i$ in an iterative fashion as shown in Algorithm 1. Because the objective is non-decreasing in each update of $\mathbf{y}$ or $\mathbf{P}$, the algorithm guarantees to converge asymptotically.

\setcounter{algocf}{0}
\begin{algorithm}
\footnotesize
 \SetAlgoLined
 \SetKwData{Left}{left}\SetKwData{This}{this}\SetKwData{Up}{up}
 \SetKwRepeat{doWhile}{do}{while}
 \SetKwFunction{Union}{Union}\SetKwFunction{FindCompress}{FindCompress}
 \SetKwInOut{Input}{input}\SetKwInOut{Output}{output}
 \textbf{initialize}: Initialize $\mathbf{P}$ to a feasible value.\\
\textbf{repeat} \\
\quad  Update $\mathbf{y}$ given $\mathbf{P}$ by (14).\\
\quad Update $\mathbf{P}$ by solving the convex
problem (12) under fixed $\mathbf{y}$.\\
\textbf{until} Convergence condition is satisfied.\\
\caption{ FP for Power Control. }
\label{alg1}
\end{algorithm}

Now that $F(\mathcal{N}_0,\alpha )$ can be efficiently obtained, we can find the optimal time allocation given $\mathcal{N}_0$, denoted by
\begin{align*}
F(\mathcal{N}_0) = \max_{\alpha\in [0,1]} F(\mathcal{N}_0,\alpha )\tag{15}
\end{align*}
via a simple one dimension search over $\alpha\in [0,1]$. One efficient method is gold section search \cite{2020:}, which is omitted here for brevity.

\begin{figure}
  \centering
   \begin{center}
      \includegraphics[width=0.35\textwidth]{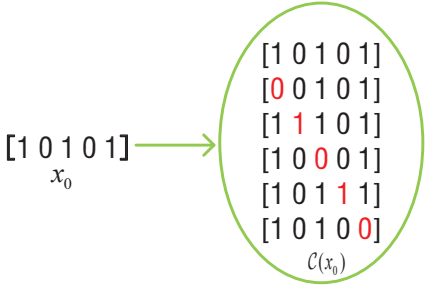}
   \end{center}
  \caption{The sampling sets $\mathcal{C}(x_0)$  of a given mode selection $x_0$}
  \label{Fig.3}
\end{figure}

\subsection{Stochastic Local Search for Mode Selection Optimization }
The remaining problem is to find the optimal binary computation mode selection vector $\mathbf{x}$. From the above discussions, we denote the achievable rate given a mode selection vector $\mathbf{x}$ as
 \begin{align*}
 F(x) = \max_{\alpha} F(x,\alpha)
\tag{16}
\end{align*}
Here, we apply a Stochastic Local Search algorithm to find the optimal $\mathbf{x}$.

The proposed  algorithm iteratively updates $\mathbf{x}$ in a probabilistic manner, while at most one of the $N$ entries of $\mathbf{x}$ is changed in each iteration.  We denote the mode selection vector in the $l$-th iteration as $\mathbf{x}^l$, where $\mathbf{x}^0$ denotes the initial mode selection solution. Then, in the $l$-th iteration, we consider $(N+1)$ candidate solutions, denoted by $\mathbf{\mathcal{C}}( \mathbf{x}^l)$,  including $\mathbf{x}^l$ itself and $N$ neighboring solutions that are off by one bit \cite{17:}. As an illustrating example, the candidate solution set of a specific $x_0$ is shown in Fig. 2. We select one of the  $(N+1)$ candidate solutions as $\mathbf{x}^{l+1}$ in the subsequent iteration.
The update is performed  according to the following probability
\begin{align*}
p(\mathbf{\mathcal{C}}(\mathbf{x}^l)_i)=\frac{\exp(-\beta/{F(\mathbf{\mathcal{C}}( \mathbf{x}^l)_i)})}{{\sum_{{\mathbf{\mathcal{C}}( \mathbf{x}^l)_i}\in{\mathbf{\mathcal{C}}( \mathbf{x}^l)}}}\exp(-\beta/{F(\mathbf{\mathcal{C}}( \mathbf{x}^l)_i)})}  ,\tag{17}
\end{align*}
 $i=1,\cdots,N+1$,  where $\mathbf{\mathcal{C}}( \mathbf{x}^l)_i$ denotes $i$-th feasible solution in candidate solution set $\mathbf{\mathcal{C}}( \mathbf{x}^l)$ and $\beta$ denotes a non-negative temperature parameter. Intuitively,  $\mathbf{\mathcal{C}}(\mathbf{x}^l)_i$ that yields a higher $F(\mathbf{\mathcal{C}}(\mathbf{x}^l)_i)$ value will be picked with a higher probability, such that (P1) tends to converge to a better solution upon each iteration. Meanwhile, it also has non-zero probability to accept solutions that are worse than the local optimum, thus it is able to escape from the ``trap" of of local optimum. Besides, by gradually increasing the temperature parameter $\beta$ as the iterations proceed, we can balance the chance of exploitation (large $\beta$) and exploration (small $\beta$) over time to converge to an optimum. The overall algorithm to solve (P1) is summarized in Algorithm 2.

\setcounter{algocf}{1}
\begin{algorithm}

\footnotesize
 \SetAlgoLined
 \SetKwData{Left}{left}\SetKwData{This}{this}\SetKwData{Up}{up}
 \SetKwRepeat{doWhile}{do}{while}
 \SetKwFunction{Union}{Union}\SetKwFunction{FindCompress}{FindCompress}
 \SetKwInOut{Input}{input}\SetKwInOut{Output}{output}
 \textbf{initialize}: Randomly generate a initial solution $\mathbf{x}^l$  and set $l=0$, initial temperature $\beta$.\\
\textbf{repeat} \\
\quad  Generate the candidate solution set $\mathbf{\mathcal{C}}( \mathbf{x}^l)$.\\

\quad  For each candidate solution $C(x^l)_i \in C(x^l)$, calculate $F(C(x^l)_i)$ in (16) by optimizing the corresponding power allocation and WPT time.\\
\quad  Calculate the probability of each solution $\mathbf{\mathcal{C}}(\mathbf{x}^l)_i$ according to (17).\\
 \quad Pick a solution at random with probability specified by (17) as the initial solution $\mathbf{x}^{l+1}$ of  the subsequent iteration.\\
\quad Set $l{\leftarrow}l+1$, $\beta \leftarrow \beta \cdot \log(1+l)$.\\

\textbf{until} the  objective value of (P1) converges.\\
\caption{ Stochastic Local Search  method for solving problem (P1)}
\label{alg1}
\end{algorithm}

\section{Simulation Results}
In this section, we  evaluate the performance of the proposed algorithm using simulations. In all simulations, we use the parameters of Powercast TX91501-3W transmitter with $P_0$ = 3W (Watt) as the energy transmitter at the HAP and those of P2110 Power harvester as the energy receiver at each WD with $\nu=0.51$ energy harvesting  efficiency. Without loss of generality, it is assumed that the noise power is set $N_0=-140$ dBm for all receivers, $q_{max}$=1mW, and $T=1$. The wireless channel gain $h_i$ between any two nodes, either HAP or a WD, follows the free-space path loss model
\begin{align*}
 h_i=G_A\left(\frac{3\times10^8}{4\pi D_i f_c}\right)^\lambda , i=1,...N,\tag{18}
\end{align*}
 where $D_i$ denotes the distance  between the HAP and $i$-th WD, $G_A=4.11$ denotes the antenna power gain, $\lambda=2.8$ denotes the path-loss exponent, and $f_c=915$ MHz denotes the carrier frequency. Besides, we set equal computing efficiency
parameter $k_i=10^{-26}, i=1,...,N,$ and $C_i=100$ for all the WDs \cite{20:}. For the data offloading mode, the bandwidth $B = 10$ MHz and the processing gain of the system $G = 128$. Without loss of generality, we assume zero residual energy of the previous time frame $e_i=0, i=1,...,N$.
\begin{figure}
  \centering
   \begin{center}
      \includegraphics[width=0.45\textwidth]{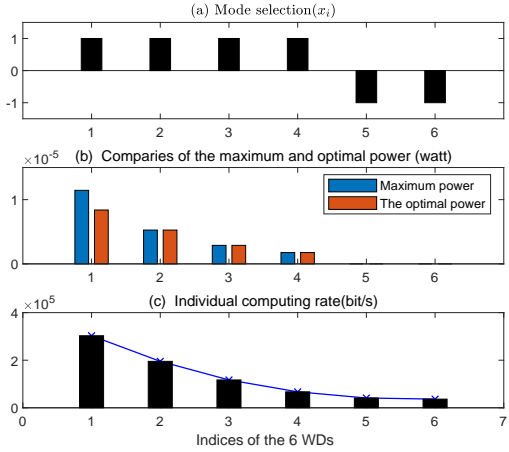}
   \end{center}

  \caption{ The figure shows the optimal computing mode solution, the comparison between the maximum power and the optimal power, and individual computing rates.}
  \label{Fig.3}
\end{figure}

In Fig. 3, we first study  some  properties of the optimal solution to (P1), where we assume a homogeneous special case with $w_i=1$ for all the WDs. For simplicity of illustration, we consider $N=6$ and $D_i=3+(i-1)$ meters,  $i=1,...,6$. Clearly, the wireless channel of WD$_1$ $(D_1=3)$ that is closest to the AP is the strongest, while the  furthest  WD$_6$ $(D_6=8)$ has the weakest channel. In all the sub-figures, the x-axis denotes the indices of the 6 WDs. The figure above  Fig. 3(a) shows the the optimal mode selection, where $x_i=1$ denotes the $i$-th WD selects mode 1 and  $x_i=-1$ denotes the $i$-th WD selects mode 0. From the figure above, we can observe that the 4  WDs with stronger wireless channels offload their tasks for edge execution while other two weak WDs  perform  computations locally. Besides, we plot in Fig. 3(b) the maximum allowable transmission power of a single user and the optimal transmission power under power control.  WD$_1$  has the highest maximum transmission power because it harvests the most energy for information transmission.
Meanwhile, the 4 users that work in mode-1 do not always transmit at the maximum power. Specifically,  WD$_1$ transmits at a lower power than the maximum allowable power, while other three users 2-4 transmit with the maximum power. Intuitively,  the one with the strongest WD-to-HAP channel (WD$_1$) may cause  strong interference to other users if transmitting at full capacity.  From Fig. 3(c), we further observe that  mobile edge computing improves the computation rate of WDs effectively. For example,  the computing rate of the mode-0 WD$_5$ is merely $12.8\%$ higher than another mode-0 WD$_6$, while the computing rate of the mode-1 WD$_4$ is $63.1\%$ higher than WD$_5$.
\begin{figure}
  \centering
   \begin{center}
      \includegraphics[width=0.4\textwidth]{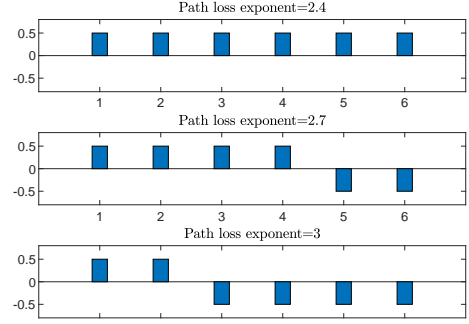}
   \end{center}

  \caption{ Optimal mode selections when path loss exponent varies.}
  \label{Fig.4}
\end{figure}

In Fig. 4, we evaluate the impact of  path loss exponent to the optimal mode selections. The three figures show the number of  mode-1  WDs  decreases when the path loss exponent $\lambda$ increases. This is because the larger channel attenuation under larger $\lambda$ leads to higher cost on task data offloading.

In Fig. 5 and 6,  we evaluate the computation performance of the proposed optimization methods. In particular,  we consider three representative benchmark methods for performance comparisons:
\begin{enumerate}
  \item
  Offloading only: all the WDs offload their tasks to the AP for remote execution;

  \item
   Local computing only: all the WDs compute their tasks locally;

  \item
   Optimal: search all the $2^{N}$ combinations of computing mode selections and output the one with the best performance.

\end{enumerate}
\begin{figure}
  \centering
   \begin{center}
      \includegraphics[width=0.4\textwidth]{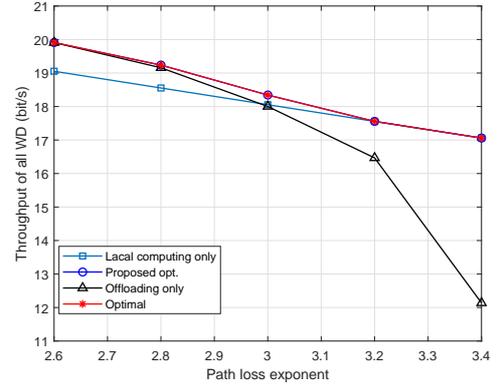}
   \end{center}

  \caption{ Computation rate  performance of different schemes when  path loss exponent varies. }
  \label{Fig.5}
\end{figure}

In Fig. 5,  we compare computation  performance of different schemes when  path loss exponent $\lambda$ increases from 2.6 to 3.4.  The network setups, e.g., the number of WDs, WD locations and weights, are the same as those used in Fig. 3. We see that the performance of all the considered schemes degrade as $\lambda$ increases (i.e., channels become weaker). Specifically,  when wireless channels are strong (that is, when path loss exponent $\lambda$ is small) the offloading only scheme achieves near optimal performance. However, as $\lambda$ increase, the performance of the offloading only scheme quickly degrades, because the offloading rates severely suffer from the weak channels. On the other hand, when $\lambda$ is large (weak channels), the all local computing scheme is optimal. The proposed optimization method achieves identical performance as the optimal scheme and significantly outperforms the other three benchmark methods.

In Fig. 6, we compare the computation performance of different methods when the number of WDs $N$ varies  from 5 to 20. For each WD$_i$, the distance to AP is uniformly generated as $D_i \sim U(2.5, 10)$, and they has homogeneous weight $w_i=1$. Each point is an average performance of $20$ independent device placement. The optimal performance is not plotted in Fig. 6, because the mode-enumeration based optimal method  is computationally infeasible when $N$ is larger, e.g., $N$=20. We compare the computation performance of the proposed optimization method  with that of the other two benchmark methods. We can see that in all the schemes, the sum rates increase  with the number of WDs. The proposed optimization method significantly outperforms the other two benchmark methods. In particular,  the average computation rate of our proposed optimization method is about 90$\%$ and 20$\%$ higher than the local-computing-only scheme and offloading-only scheme, respectively.

 In Fig. 7, we characterize the computation complexity of the proposed Stochastic Local Search algorithm. We plot the average number of  searches when the number of WDs $N$ varies from 5 to 20. Here, we consider the algorithm converges if the difference of objective value between two adjacent iterations is less than  $10^{-4}$. We see that average number of Stochastic Local Search algorithm iterations increases almost linearly  with the number of WDs. The results indicate that Stochastic Local Search algorithm can quickly converge to the global optimal solutions and is applicable to large-size networks.
\begin{figure}
  \centering
   \begin{center}
      \includegraphics[width=0.4\textwidth]{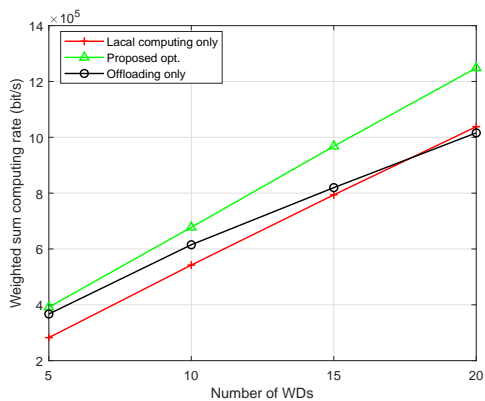}
   \end{center}

  \caption{When the number of users changes, the performance of different schemes is compared.}
  \label{Fig.6}
\end{figure}

\begin{figure}
  \centering
   \begin{center}
      \includegraphics[width=0.35\textwidth]{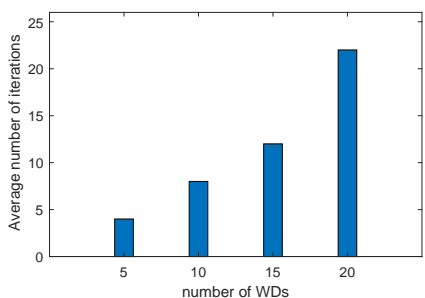}
   \end{center}

  \caption{ Average number of Stochastic Local Search algorithm search times when the number of WDs varies.}
  \label{Fig.7}
\end{figure}

\section{Conclusions}
In this paper, we studied the resource allocation problem of a multi-user wireless powered MEC system using DS-CDMA to coordinate the task offloading of the users. We formulated the computation rate maximization problem as a joint optimization of individual computing mode selection, transmission power control and WPT time allocation. To tackle the problem, we first assumed that  the computation mode selection is given and applied  a direct FP approach to solve the rate maximization problem in the uplink interference channel. On top of that, a simple linear search method is applied to find the optimal WPT time.  In addition, we further proposed a  Stochastic Local Search algorithm to optimize the individual computing mode selections. By comparing with representative benchmark methods, we showed that the proposed the scheme can significantly improve the computation  performance of the wireless powered MEC system based on DS-CDMA.

\end{document}